\documentclass[conference]{IEEEtran}
\usepackage{subfigure}
\usepackage{caption}
\usepackage{graphicx}
\usepackage[lined,boxed,commentsnumbered,ruled]{algorithm2e}
\usepackage{amsmath}
\usepackage{amsfonts}
\usepackage{verbatim}
\usepackage{array}
\usepackage{comment}
\usepackage{balance}
\makeatother                
\begin{document}

\title{VBCA: A Virtual Forces Clustering Algorithm for Autonomous Aerial Drone Systems}

\author{Matthias R. Brust$^1$, Mustafa \.{I}lhan Akba\c{s}$^2$ and Damla Turgut$^2$\\
$^1$Singapore University of Technology and Design, Singapore\\
$^2$University of Central Florida, USA\\

matthias\_brust@sutd.edu.sg, miakbas@eecs.ucf.edu, turgut@eecs.ucf.edu}

\maketitle
\begin{abstract}
We consider the positioning problem of aerial drone systems for efficient three-dimensional (3-D) coverage. Our solution draws from molecular geometry, where forces among electron pairs surrounding a central atom arrange their positions. 

In this paper, we propose a 3-D clustering algorithm for autonomous positioning (VBCA) of aerial drone networks based on virtual forces. These virtual forces induce interactions among drones and structure the system topology. The advantages of our approach are that (1) virtual forces enable drones to self-organize the positioning process and (2) VBCA can be implemented entirely localized.

Extensive simulations show that our virtual forces clustering approach produces scalable 3-D topologies exhibiting near-optimal volume coverage. VBCA triggers efficient topology rearrangement for an altering number of nodes, while providing network connectivity to the central drone. We also draw a comparison of volume coverage achieved by VBCA against existing approaches and find VBCA up to 40\% more efficient. 

\end{abstract}

\IEEEpeerreviewmaketitle

\section{Introduction}
\label{Introduction}

Many aerial drones have integrated wireless networking adaptors to enable the establishment and deployment of interconnected autonomously acting drone systems \cite{Akbas11a}. This flying ad hoc network (FANET) can scan its environment and react in real-time to adjust location and formation depending on mission requirements \cite{Bekmezci13,Brust15b}. Evidently, these capabilities provide tremendous potential for FANETs in applications such as surveillance, search and rescue operations in disaster recovery, and target localization \cite{Brust15}. 

In this paper, we introduce a virtual forces based clustering algorithm (VBCA) for three-dimensional (3-D) positioning of drones for efficient volume coverage, while maintaining a fully connected communication network. The resulting network topology after the drones' positioning is a critical aspect to achieve this goal. \cite{Brust15}.

We use a localized virtual forces approach to create the interactions among drones and to form the topology of the system. The virtual forces are drawn from the Valence Shell Electron Pair Repulsion (VSEPR) model \cite{Gillespie57}, where forces among electron pairs surrounding a central atom actively position the entities of a system.


The key goal of VBCA is providing the same efficient VSEPR molecular geometries in the setting of a FANET, while providing scalability analogously to the 3-D geometries of VSEPR model. Each drone's position is determined by the distance and role of its neighboring drones. The central drone acts as a connector influencing the entire topology of the network geometry while individual drones are only affected by their direct neighbors.

Simulation results show that VBCA ensures resilience of the drone network resulting from self-organizing properties of the proposed clustering process to a changing number of drones (for scalability in clustering processes cf.  \cite{brust2007adaptive, brust2007waca, brust2008dynamic}). VBCA is computationally efficient while provideing network-wide connectivity. An altering number of nodes causes VBCA to trigger efficient topological rearrangement proofing its adaptivity capabilities for real-world applications. We also draw a comparison to the volume coverage achieved by VBCA against existing approaches and find VBCA up to 40\% more efficient. 

The remainder of the paper is organized as follows. A detailed description of our approach is given in Section II. We present and discuss the simulation results in Section III. Section IV summarizes the related work, and we conclude in Section V.

\begin{figure*}
  \centering

\includegraphics[width=7in]{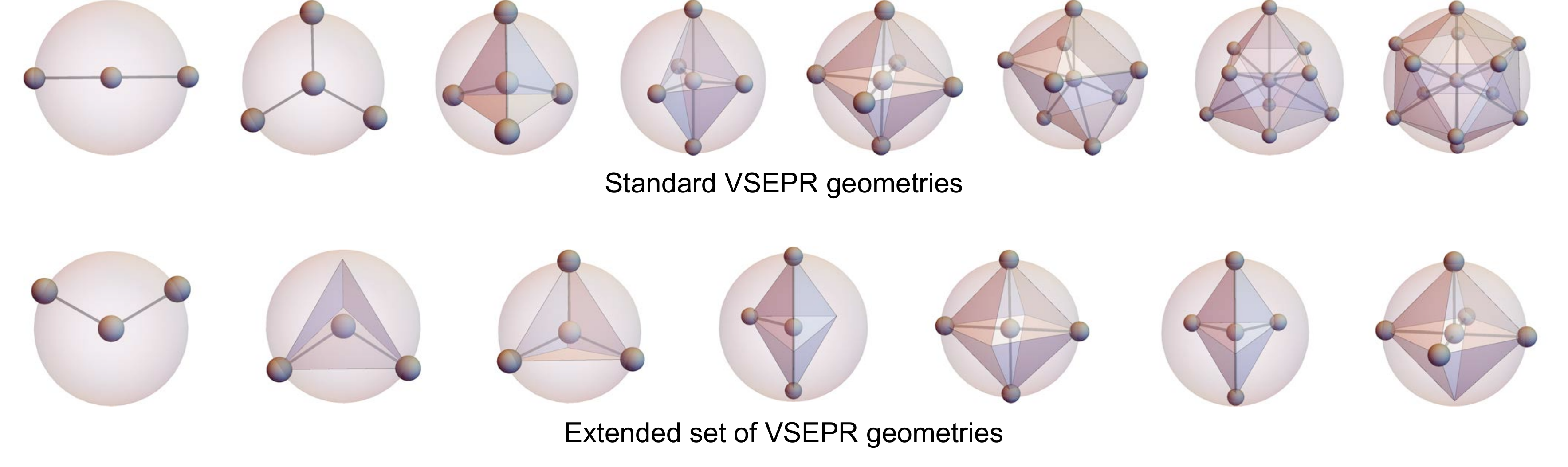}

\caption{The VSEPR model describes standard geometries as well as supplemental solutions for certain system configurations.}
\label{fig:VSEPRgeometries}
\end{figure*}

\section{A VSEPR Forces based Clustering Algorithm}
\label{OurProtocol}

We approach the described 3-D drone positioning problem with a method inspired by the VSEPR model \cite{Gillespie57}. 
The purpose of the VSEPR model is to predict the geometric formations of molecules by using the peripheral atom alignments around a central atom as illustrated in Fig. \ref{fig:VSEPRgeometries}.


The geometric formations of VSEPR model are based on the notion that electron pairs arrange themselves to be as far apart as possible from each other with minimal repulsion. The arrangement of distances within the VSEPR model is utilized in VBCA to maximize the total covered volume by the drones. Hence, the goal of our approach is the determination of drone system topology by using the VSEPR model. The VSEPR model geometries fulfill our optimization requirements by (1) providing an optimal covered volume and (2) maintaining a fully connected communication network.


\subsection{System model}

The drone system $N$ considered in this paper is composed of $|N|$ nodes. Each drone $n \in N$ is capable to communicate with neighboring drones in the communication range $r$ and collecting data from the environment. We define a central node $c$, which is assumed to be more capable compared to the remaining nodes in terms of energy resources, sensor receptors and communication adaptors. The central drone with its stronger capabilities can be used as an interface to a wider communication backbone.

The initialization of the clustering process in VBCA starts with a role assignment for the central drone, which is valid throughout the lifetime of this particular system. More specifically, the central drone is initialized with its role and all other drones are assigned as the peripheral drones.
The peripheral nodes are homogeneous regarding communication ranges, and each drone communicates exclusively with its direct neighbors. According to this procedure, each drone must have the central drone in its transmission range for the network connectivity.

The high mobility of the surveilled objects and the frequent changes in the environment are important examples of the challenges faced in drone systems. Since these conditions can result in malfunctioning drones, the possibility of a change in the number of drones must be taken into account. The proposed approach deals with these challenges in real-time by arranging the topology according to the total number of drones in the system. 

\subsection{VBCA protocol}

The VSEPR model describes the molecular geometry formation process by a mutual repulsion process by negative electrons, which causes self-arrangement of the electron pairs towards a steady-state with minimal repulsive forces acting on the electrons.
Therefore, we leverage the VSEPR forces as virtual forces acting between the nodes in a drone ad hoc network. These forces are defined as follows:

\begin{enumerate}
\item Attraction: The \emph{attraction force} acts between the central drone and each of the remaining drones. Therefore, the central drone acts as an attracting source for the peripheral drones. This attracting force pulls the drones to the center of the topology to maintain network connectivity during the operation. 
\item Repulsion: The \emph{repulsion force} acts among all drones except the central drone. The repulsion forces refrain the drone system to collapse toward the center as a result of the attraction force. Therefore, the distances to the central drone increase as the drones repel each other, until the forces acting on the nodes are balanced.
\end{enumerate}

Algorithm \ref{algorithm1} presents the utilization of the virtual forces and their collaborative function on the drone network. 

\title{Algorithm}
\maketitle
\begin{algorithm}
\LinesNumbered 
\SetKwInOut{Input}{Input}
\SetKwInOut{Output}{Output} 
\Input{A set of drones $D$ with its positions $p$ and velocities $V$} 
\Output{A set of drones $D$ with adjusted positions $p$ and velocities $V$} 
\BlankLine
$cNode\leftarrow\mbox{central node}$\; 
$pNode\leftarrow\mbox{pheripheral node}$\;
\ForEach{$d\in D$}{ 
   $N(d)$ $\leftarrow$ FindDirectNeighbors($d$)\; 
   \ForEach{$n\in N$}{ 
      \If{$Type\left(d\right)=cNode\,\mbox{\&}\,Type\left(n\right)=pNode$}{
         $V_{attract,d}\left(t\right)\leftarrow\left[a*\left(\frac{\sum_{n\in N}p_{n}\left(t-1\right)}{|N|}\right)-p_{d}\left(t-1\right)\right]$\;
      }
      \If{$Type\left(d\right)=pNode\,\mbox{\&}\,Type\left(n\right)=pNode$}{
         $V_{repuls,d}\left(t\right)$$\leftarrow$ ${\scriptstyle \left[r*\sum_{d}\left(\frac{1}{\delta\left(x_{d,n}\left(t-1\right)\right)},\frac{1}{\delta\left(y_{d,n}\left(t-1\right)\right)},\frac{1}{\delta\left(z_{d,n}\left(t-1\right)\right)}\right)\right]}$\;
      }
   }
   $V_{d}\left(t\right)\leftarrow\left[V_{attraction,d}\left(t\right)+V_{repulsion,d}\left(t\right)\right]$\;
   $p_{d}\left(t\right)\leftarrow\left[p_{d}\left(t-1\right)+V_{d}\left(t\right)\right]$\;
} 
\caption{VBCA}\label{algorithm1}
\end{algorithm}

The resultant velocity vector $V_{d}$ for drone $d$ after time $t$ is calculated in Algorithm \ref{algorithm1} by considering the combinational effects of the attraction forces $V_{attract,d}\left(t\right)$ and the repulsion forces $V_{repuls,d}\left(t\right)$. $V_{attract,d}\left(t\right)$ is the velocity vector for the attraction force of drone $d$ at time $t$ and $V_{repuls,d}\left(t\right)$ is the velocity vector for the repulsion force of drone $d$ at time $t$. The current position of the drone $d$ is calculated by the velocity vector $\left[p_{d}\left(t-1\right)+V_{d}\left(t\right)\right]$.

The method $Type(d)$ returns the drone type of $d$ (i.e., whether it is a central drone or a peripheral drone) and $N\left(d\right)$ is the set of neighboring drones which are directly connected to $d$. For this, we define the function $\delta\left(x_{d,n}\left(t\right)\right)$ to measure the dimensional distance between two drones $d$ and $n$. According to this function, the distance in direction $x$ at $t$ is given as follows:

\begin{equation}
    \delta\left(x_{d,n}\left(t\right)\right)=p\left(x_{d}(t)\right)-p\left(x_{n}(t)\right)\nonumber\\
\end{equation}

Similar equations are used for $y$ and $z$ directions, whereby $p\left(d\left(t\right)\right)$ is the position of the current drone at time $t$ and $p\left(n\left(t\right)\right)$ is the position of a neighbor from set $N$ at time $t$. The parameter $a$ and $r$ are introduced to calibrate the strength of the attraction force and repulsion force, respectively.

It is necessary to adjust the absolute strength level of each force such that the resulting topologies resemble the geometries described by the VSEPR model. For this reason, we use specific parameters in VBCA, which balance the strength of the attraction and repulsion forces (see Algorithm \ref{algorithm1}). This fundamental formulation of two forces and their usage in the algorithm resembles the idea in the geometries of the VSEPR model. 

The application of VBCA results in a self-stabilizing system with VSEPR model-like topologies as steady-state. Hence, the formation and maintenance of these topologies are critical for a successful execution of VBCA. Locations of the drones must be near-optimal in terms of closeness to the central drone and their relative positions according to the other peripheral nodes to obtain steady-state topologies. For this purpose, a network parameter ($CP$) is introduced following, and all possible topologies formed by the virtual forces in our approach are considered: 

\subsubsection{Compactness parameter}

The transmission range $r$ is an influence on the distances used in the topologies determined by our model. However, the \emph{virtual} forces act on each drone in different directions depending on its specific position in the topology. Therefore, the distances that the drones communicate in the steady-state of the system are different than $r$. The distances in VSEPR model are determined by physical rules. However in VBCA, the topology of the aerial network can be preserved while adjusting the distances according to the application conditions.    

We introduce the \emph{compactness parameter} $CP$ to regulate 
the closeness of drones in the geometric formation. $CP$ presents the option to manipulate how compact the steady-state geometry must be. This is an advantage of VBCA as it eliminates the requirement for constant distance values, while having no specific influence on the considered geometry.
Therefore, $CP$ provides the capability to dynamically change distances in the topology for scenario-specific requirements such as following the contours of the observed environment.

\subsubsection{Extended set of VSEPR topologies}

There are suplemental cases under the VSEPR model, for which multiple extra topologies are possible for an identical number number of atoms. 
This extended set of topologies results from multiple geometrical solutions possible by the forces-dynamics in the system. This solution space is built in by the heterogeneity property of the model. Heterogeneity in VBCA plays a crucial role in two aspects: (a) the force types are different (such as attraction and repulsion) and (b) the strength/magnitude of each force type is alterable. 

The connectivity constraints imposed by our scenario keep the drones within a certain distance to each other. In other words, the peripheral drones can achieve better volume coverage unless they have the central drone connectivity constraint. There are extensions for the VSEPR model to deal with the interconnection of multiple topologies formed by multiple central drones \cite{Akbas12}. In this paper, however, we focus on the cases dealing with a single central node and reserve multiple central node scenarios for future work.

\section{Simulation Study}
\label{SimulationStudy}

VBCA has been implemented in Wolfram Mathematica for system visualization as well as performance evaluation. Simulation results are measured when the system reaches a steady-state. The condition for steady-state is defined as a minimum movement threshold, while the entire system remains fully connected.

\subsection{Simulation settings}

We focus on two aspects for the performance evaluation of VBCA: The effective volume coverage of the resulting topologies and the average distance from the central node to the peripheral nodes and its variations. 

The compactness parameter $CP$ is used to define the network density and its value varies from 10 to 70 with an increment of 10 in the simulations. We set the transmission range to $r_t = 80 m$, while the collision range $r_c$ is set to $r_c=60 m$. The number of drones in a cluster varies between 3 to 10 according to the number used in the specific simulations. 

\subsection{Simulation execution}

For each simulation, all nodes are initially deployed at the same coordinates with one dedicated node as central node. After VBCA has been initialized, the virtual forces start to impact the topological features of the drone cluster and the topology unfolds toward its steady-state configuration. VBCA leads to an autonomous system of drones for which the steady-state position of the drones are optimal in terms of closeness to the central drone and the repulsion among each other. 

Our simulations reveal that the VBCA clustered topologies in their steady-state configuration are exclusively VSEPR geometries from the standard set as well as from the extended set of VSEPR geometries (cf. Fig. \ref{fig:VSEPRgeometries}).

\begin{figure*}
  \centering
\includegraphics[width=7.0in]{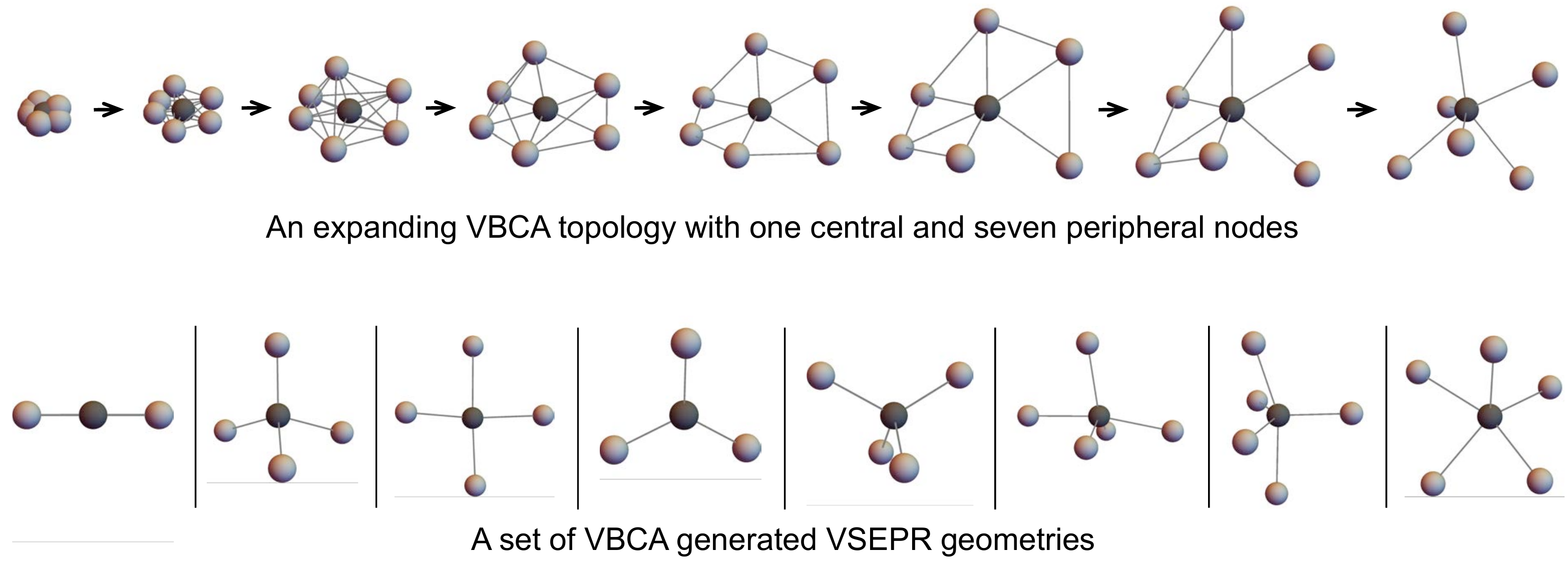}
\caption{Expanding VBCA topology reproducing VSEPR geometries and a set of VBCA generated VSEPR geometries.}
\label{fig:ExampleExpand}
\end{figure*}

Fig. \ref{fig:ExampleExpand} demonstrates how a VBCA geometry unfolds on the case of a system with one central and seven peripheral nodes. Further simulations conducted with VBCA also reproduced successfully an extensive set of VSEPR geometries, including standard VSEPR and extended VSEPR geometries. Some results are visually shown in Fig. \ref{fig:ExampleExpand}.

\subsection{Simulation results}

\subsubsection{Compactness parameter vs. node distance}

\begin{figure}
\centering
\includegraphics[width=3.4in,height=2.4in]{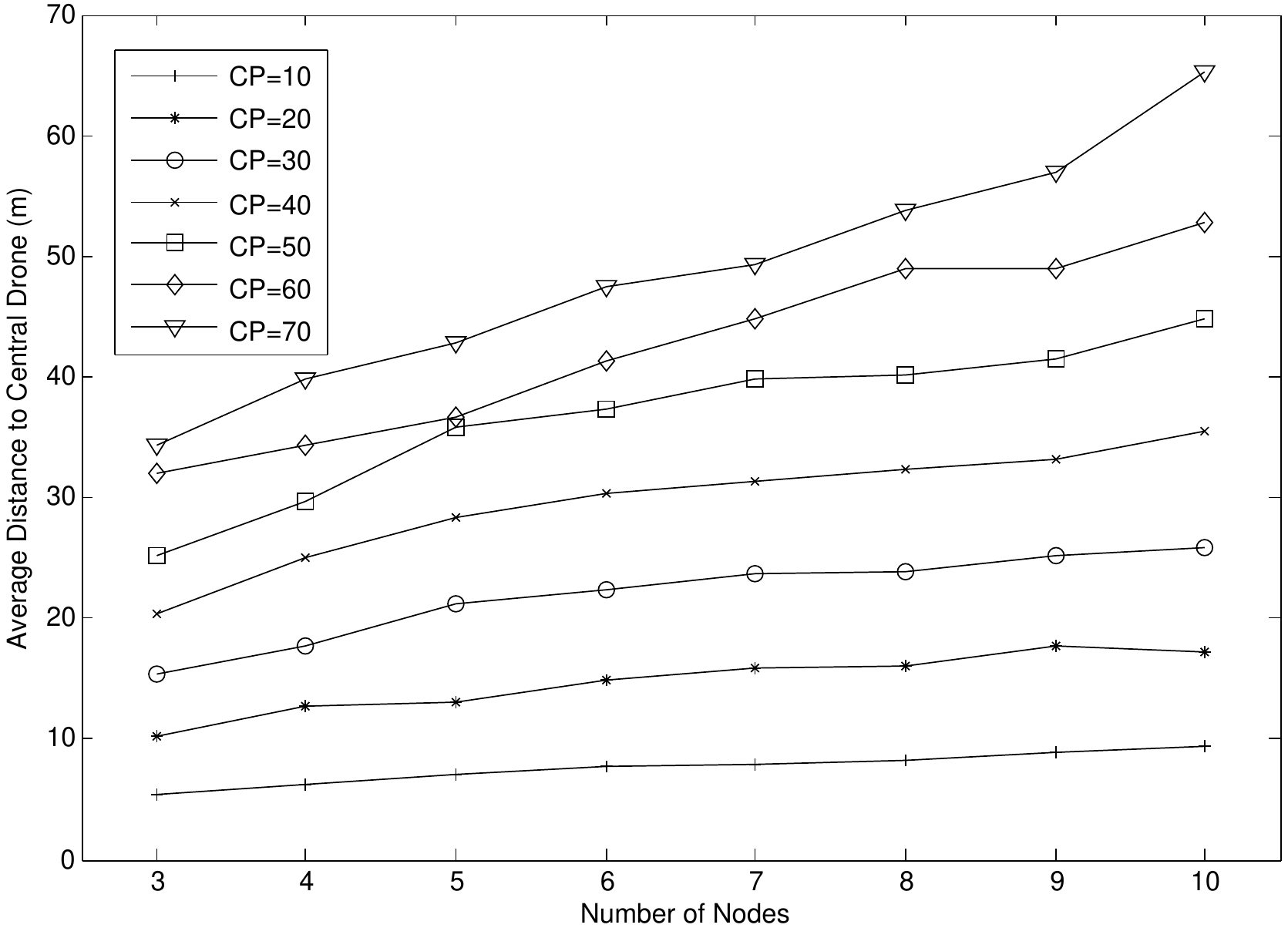}
\caption{Average distance from each node to the central node for different $CP$ values.}
\label{fig:Results1}
\end{figure}

In Fig. \ref{fig:Results1}, we demonstrate the relation between the value of $CP$ and the average distance from the peripheral nodes to the central node for topologies with different number of nodes. 

Fig. \ref{fig:Results1} shows the effect of the same $CP$ value for different number of nodes on the variance of the average distance. For instance, when $CP$ is 10, the average distance to the central node varies between $5$ to $8 m$, in a range of 3. However, when $CP$ is 70, this range increases $30 m$. Therefore, for the scenarios with larger $CP$ values, the drones must be able to exchange data with the central node from further distances as the number of drones increases in the network.

The results also show the effect of the $CP$ in average distance from the central node to the peripheral node. As the $CP$ becomes larger, the average distance to the central node increases its values for the same number of nodes. For topologies with a larger number of nodes, the increasing $CP$ values have a more significant effect on the average distance to the central node. 

The results in Fig. \ref{fig:Results1} show that the transmission range $r_t$ and the $CP$ create important requirements for the selection of each other. 
For instance, when the drones have a relatively small transmission range, the $CP$ cannot be chosen to be very large as the drones operate in short distance to each other. 

\subsubsection{Stability of the topologies}

\begin{figure}
\centering
\includegraphics[width=3.5in,height=2.5in]{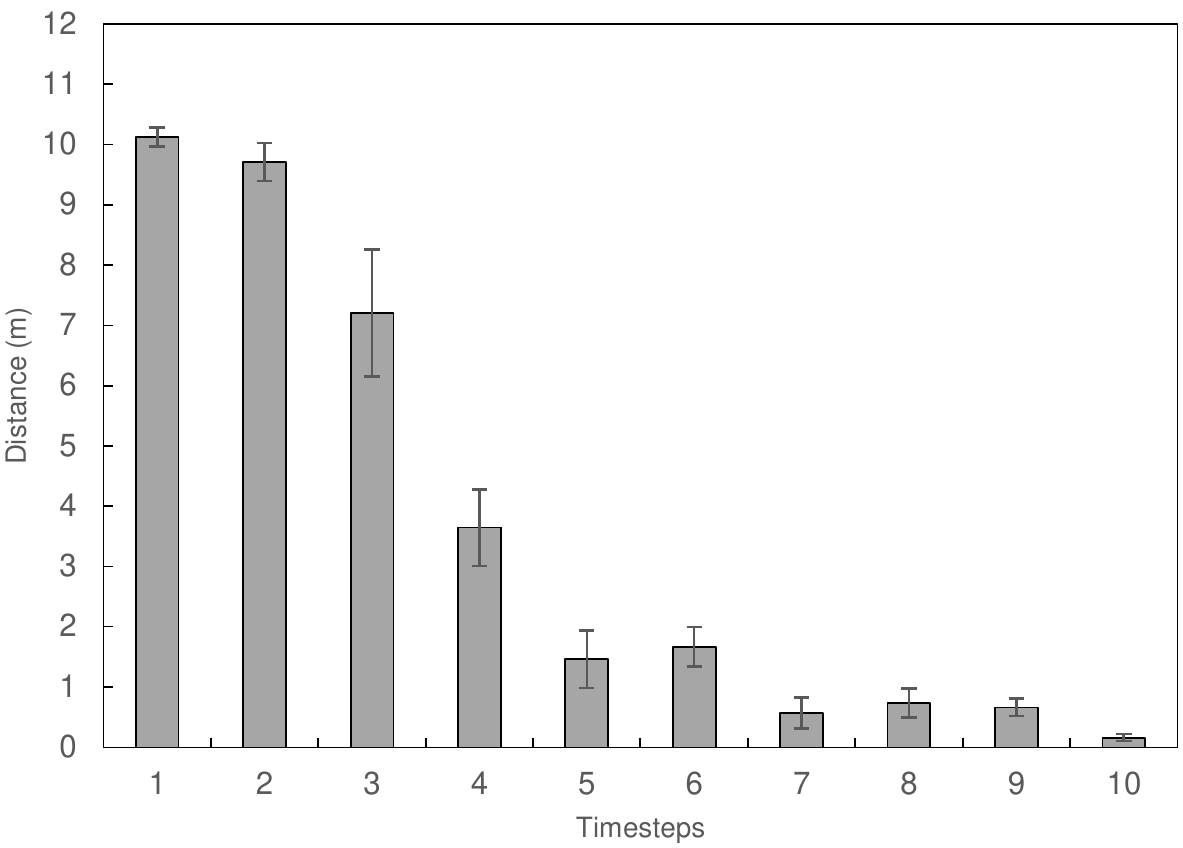}
\caption{The variation in the position of each node in reference to the central node.}
\label{fig:Results5}
\end{figure}

In Fig. \ref{fig:Results5}, the average position variation of each node is demonstrated for an eight drone scenario. Since the position of the central node does not change over time, it is not included 
and the position variations of other nodes are absolute values. Hence, whether the node approaches  the central node or drifts away, the absolute value of change in the distance is given. 

The results in Fig. \ref{fig:Results5} show how the drone topology becomes stable after the self organization period. The variation in the positions of the nodes asymptotically converges to $0.15 m$ after 10 time steps. The stability of the topology is critical for the efficient operation of the drone network since it provides optimum volume coverage for the lifetime of the scenario. 


\subsubsection{Volume coverage for different topologies}

The main purpose of VBCA is to provide an efficient positioning in terms of maximizing the total covered volume around the central drone. 

The inputs for calculating the volume coverage are the number of spherical covered volumes, the positions and the communication ranges of drones. Hence the theoretical maximum volume coverage  $V_m$ for $n$ drones is calculated as follows:

\begin{equation}
	V_m = \sum\limits_{i=1}^n \frac{4}{3} \pi {c_i}^3, \nonumber \\
\end{equation}

\noindent where $c_i$ is the observation range of the drone $i$.  

The total volume coverage of the drones is smaller than $V_m$ when there are intersections of the coverage volumes in the scenario. Then, the total volume coverage is calculated by using each drone's individual volume coverage and taking the union of this result considering the drones' respective coordinates. 

\begin{figure}
\centering
\includegraphics[width=3.4in,height=2.4in]{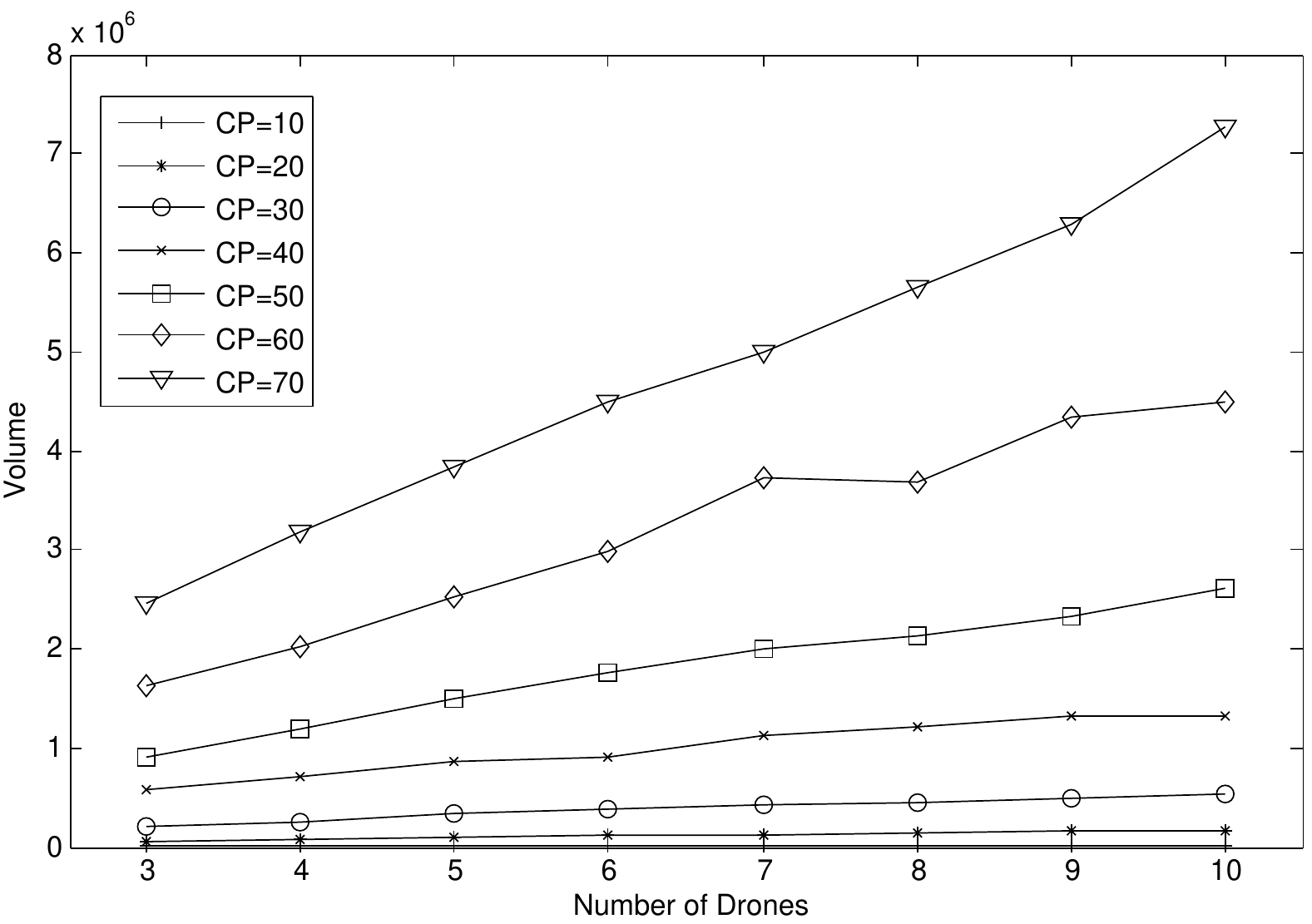}
\caption{Comparison of total coverage volume of geometry configuration with different number of nodes.}
\label{fig:Results2}
\end{figure}

Fig. \ref{fig:Results2} shows the drone cluster coverage values for topologies with one central drone and with an increased number of drones. For all $CP$ values, the total volume coverage follows a linear increase as the total number of drones increases from 3 to 10. However, the rate of this increase varies for different $CP$ values.

\subsubsection{Volume coverage comparison}

\begin{figure}
\centering
\includegraphics[width=3.4in,height=2.4in]{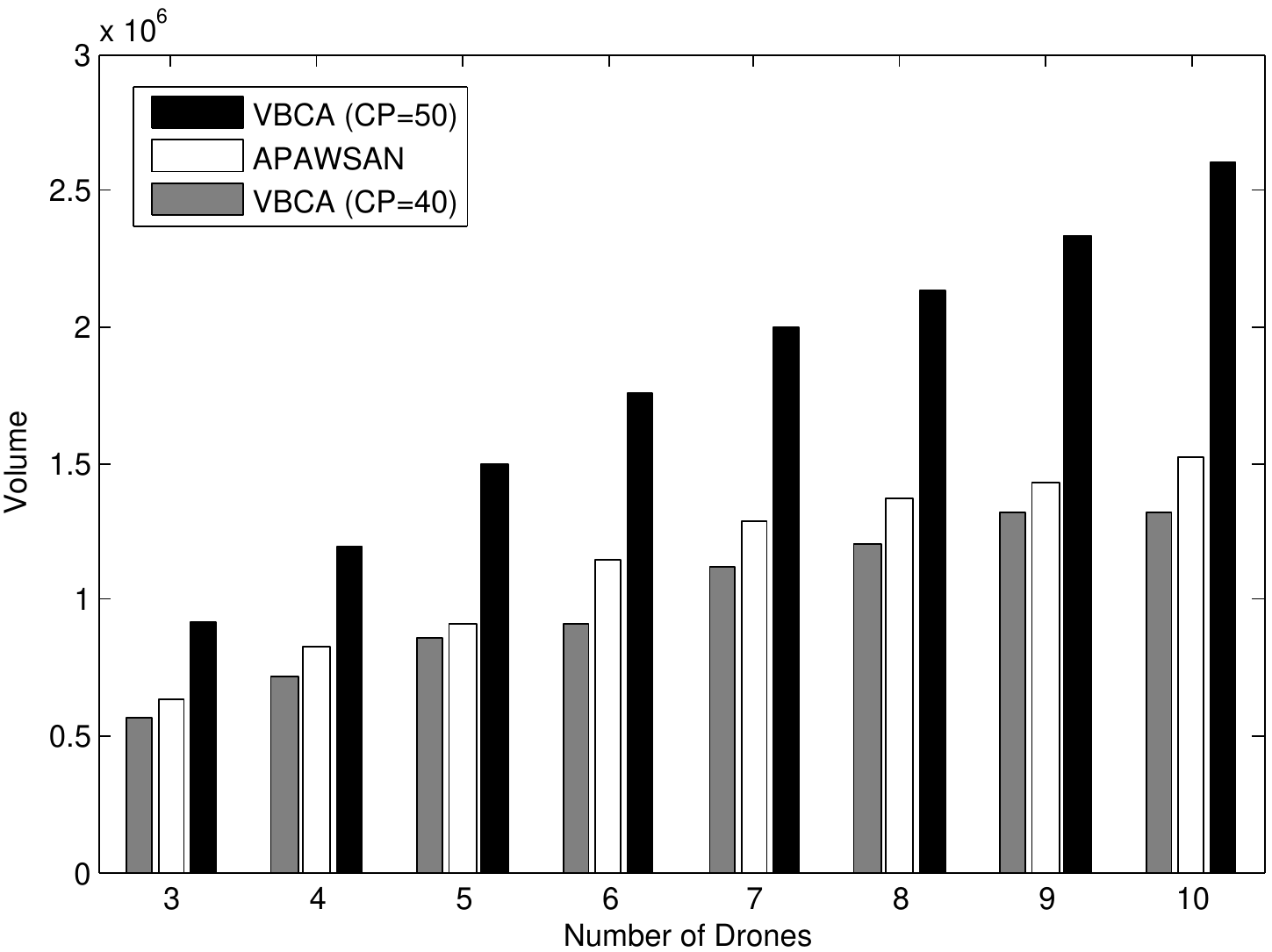}
\caption{Volume coverage of VBCA compared to volume coverage of APAWSAN.}
\label{fig:Results3}
\end{figure}

We also obtained results for comparing VBCA to APAWSAN \cite{Akbas11a} and the 3-D deployment approach by Lee et al. \cite{Lee10} regarding efficiency of total volume coverage. APAWSAN provides 1-hop connectivity from the central drone to the peripheral drones by placing each of these drones in the exact positions calculated according to VSEPR model. In contrast, VBCA uses a force based approach producing these topologies instead of executing calculations for exact positions. Therefore, it is important to see how this force based method affects the individual node positions and the covered volume. Fig. \ref{fig:Results3} shows results for VBCA with $CP$ set to 40 and 50, and APAWSAN with a drone transmission range of $40 m$.

\section{Related Work}
\label{RelatedWork}


There are various approaches for the 3-D coordination and positioning of aerial networks. The aerial network presented by Elston and Frew \cite{Elston08} contains a central ship with multiple drones, which use field tracking for the hierarchy. Dumiak \cite{Dumiak09} proposes a coordination mechanism for aerial networks to complete the tasks by multiple drones. 
UAVNet is an autonomous deployment framework for FANETs by Morgenthaler et al. \cite{Morgenthaler12}. UAVNet aims to provide an efficient way to construct a communication network, which can be controlled by a single remote user. 
The mobility prediction clustering algorithm \cite{Zang11} uses the dictionary tree structure prediction algorithm with link expiration time mobility model to overcome the challenge of frequent cluster updates by predicting the network topology updates. Luo et al. \cite{Luo13} propose a positioning and collision avoidance strategy for UAVs in search scenarios, which uses received signal strength (RSS) from the onboard communication module. De Medeiros et al. \cite{Paixao14} propose a prognostics and health monitoring based multi-UAV task assignment approach to include system probability of failure into task assignment in a drone system. This method assigns tasks based on the drone health condition using the Receding Horizon Task Assignment (RHTA) algorithm. 

Brust and Strimbu \cite{Brust15} introduce a UAV networked swarm model for forestry assessment and environmental monitoring. The UAV swarm is able to establish, maintain multi-hop connectivity and avoid obstacles, while assessing the forest environment (e.g. tree localization, tree mapping). Iskandarani et al. \cite{Iskandarani14} use Linear Model Predictive Control (LMPC) to implement line abreast, triangular and cross formation flights for drones in simulations and experiments. The Reconfigurable Flight Control System Architecture (RFCSA) \cite{Deng12} is a control framework for small UAVs. RFCSA utilizes a different module for each function of a drone to minimize the complexity of implementation and coordination during the flight.

A geometry-based deployment and positioning strategy is used in VBCA. Lee et al. \cite{Lee10} proposes a geometric approach for addressing the deployment of an autonomous mobile robot swarm randomly distributed in 3-D space. Through selective and dynamic interaction, four robots form a tetrahedron topology. The Regular Tetrahedron Formation (RTF) strategy \cite{Zeng11} by Zeng and Li is proposed for a swarm of robots which is based on a virtual spring mechanism to form the topologies. The movements at each time step are dependent on the local position information of three neighbors. 

In an alternative geometric approach, Akbas et al. \cite{Akbas11a, Li14} use VSEPR model \cite{Gillespie57} to present a node positioning strategy, APAWSAN, for drone networks. According to VSEPR model, the peripheral atom locations in a molecule are defined by the repulsion forces among the electron pairs. 
APAWSAN utilizes the VSEPR model to position the drones with actor roles around one central drone, which has the role of a sink in a wireless sensor and actuator network. 
The other drones use the information shared by the central drone to position themselves around the central node with high network connectivity and volume coverage. 
APAWSAN calculates and provides accurate drone positions with central control, i.e. assuming centralized computation of the drones' position. However, VBCA further reduces complexity compared to APAWSAN by using a force based approach at each drone depending on local information. 

\balance

\section{Conclusion}
\label{Conclusion}



VBCA describes an autonomous and dynamic system of drones for which the steady-state location of the peripheral drones are optimal in terms of their distance to the central drone and the repulsion among each other. The resulting system provides a constraint-based efficient volume coverage with the constraint that the peripheral drones must be connected to the central atom. Thus, VBCA produces clustered topologies which accurately reflect those of the VSEPR model geometries.

Simulation results show that VBCA provides topologies with network connectivity and efficient volume coverage. Results also reveal an improved convergence of drone clusters towards a steady-state. Additionally, we show that VBCA outperforms two geometric deployment approaches from the literature in terms of volume coverage. We conclude that the nature-inspired topologies found in the molecular geometries can be created by a rather small set of computationally efficient local virtual forces to efficiently position autonomous drones.

Future work will include the implementation of VBCA topologies that handle multiple central drones as well as a protocol for the interconnection of multiple clusters. These configurations, eventually equipped with an integrated routing scheme to manage data collection and aggregation, will enable VBCA to be applied to irregular real-world objects such as bridges and buildings.

\bibliographystyle{IEEEtran}
\bibliography{VBCA}

\end{document}